\documentstyle[12pt]{article}

\begin{document}

\begin{titlepage}
\null\vspace{-62pt}

\pagestyle{empty}
\begin{center}

\vspace{1.0truein} {\Large\bf Instanton interaction in de~Sitter spacetime}

\vspace{1in}
{\large Dimitrios Metaxas} \\
\vskip .4in
{\it Department of Physics,\\
National Technical University of Athens,\\
Zografou Campus, 15780 Athens, Greece\\
metaxas@central.ntua.gr}\\

\vspace{.5in}
\centerline{\bf Abstract}

\baselineskip 18pt
\end{center}
Because of the presence of a cosmological horizon the dilute instanton gas approximation used for the derivation of the Coleman-De~Luccia tunneling rate in de~Sitter spacetime receives additional contributions due to the finite instanton separation.
Here I calculate the first corrections to the vacuum decay rate that arise from this effect and depend on the parameters of the theory and the cosmological constant of the background spacetime.

\end{titlepage}
\newpage
\pagestyle{plain}
\setcounter{page}{1}
\newpage

\section{Introduction}

The calculation of the decay rate of metastable vacua in flat and curved space-time \cite{coleman1, cdl}  and zero or finite temperature \cite{linde1} is of great interest due to various applications to problems at the interface of particle physics and cosmology. For example, the discovery of a multitude of string vacua suggests that a detailed understanding of these rates for different parameters of the models may be relevant to the investigation of the various cosmological phase transitions that took place in the evolution of the universe as we know it \cite{l1}. In particular, it is important to have quantitative results for the
progress of the phase transitions and the
 dependence of the decay rate on physical parameters such as the temperature \cite{gleiser1} and the Hubble expansion rate or the curvature of the background space-time \cite{ejw1, k1}.

To this effect, the usual semiclassical calculation at zero temperature in flat space-time, which I review in Sec.~2, needs various modifications and admits additional contributions depending on the physical situation at hand.

In Sec.~3, I give some estimates for the curved space-time contributions to the expression for the decay rate compared 
to the flat space-time result, in order to further describe the approximations used here and the motivation for this work.

In Sec.~4, I proceed to examine the corrections that arise in de~Sitter spacetime because of the presence of a cosmological horizon. The dilute instanton gas approximation then receives corrections, the first of which arises from the two-instanton interaction, and I calculate it in the thin wall and fixed background approximation. The fact that the instanton interaction in de~Sitter space-time turns out to be repulsive, essentially because of the cosmological expansion, and unlike most instanton and soliton interactions in scalar field theories, makes the calculation feasible, without the need of additional cutoffs.
The corrections obtained here are, of course, subleading with respect to the semiclassical exponential factor, they are of interest, however, because of their dependence on the various parameters of the theory, especially the cosmological constant of the background space-time.

 Finally, in Sec.~5, I conclude with some comments regarding the limitations and possible extensions of this work.

\section{Review of the flat space-time results}

In order to calculate the false vacuum decay rate
for a scalar field theory in flat, $3+1$-dimensional space-time and zero temperature one evaluates the imaginary part of the energy of the false vacuum, obtained from the path integral
\begin{equation}
Z(T) = \int [d\phi] \exp(-S_E(\phi)),
\label{z1}
\end{equation}
with the Euclidean action
\begin{equation}
S_E(\phi) =\int d^3x \int_{-T/2}^{T/2} dt_E \, L_E,
\label{flataction}
\end{equation}
and
\begin{equation}
L_E=\frac{1}{2}(\partial_{\mu}\phi)^2 + U(\phi),
\end{equation}
with $x_{\mu}=(t_E,\vec{x})$ the Euclidean space-time coordinates.
 The potential $U(\phi)$ has a metastable minimum at $\phi=\phi_{f}$ and an absolute minimum at
$\phi=\phi_{t}$, and the path integral is restricted to paths satisfying $\phi(-T/2)=\phi(T/2)=\phi_{f}$.

Then, in the limit $T\rightarrow\infty$,
\begin{equation}
Z(T)\approx\exp(-E_{f} T\Omega),
\end{equation}
where $\Omega$ is the volume of space and $E_{f}$ the energy density of the false vacuum and the imaginary part of this quantity will give the decay rate
\begin{equation}
\Gamma = -2 \frac{{\rm Im} \ln Z(T)}{T\Omega},
\label{gamma}
\end{equation}
that is, in this case, the bubble nucleation rate, or the number of true vacuum bubbles produced per unit volume per unit time.

The path integral can be evaluated in the saddle-point approximation by obtaining an instanton (bounce) solution, $\phi_b$,
of the Euclidean field equation
\begin{equation}
\Box_E\phi =\frac{dU}{d\phi},
\label{flatbounce}
\end{equation}
which, assuming $O(4)$ symmetry, becomes
\begin{equation}
\ddot{\phi}+\frac{3}{r}\,\dot{\phi}=\frac{dU}{d\phi},
\label{o4flat}
\end{equation}
with $r=\sqrt{x_{\mu}^2}$ the Euclidean radial distance.
Then
\begin{equation}
Z(T)\approx Z(\phi_f) + Z(\phi_b) + Z_2 + \ldots
\label{gas1}
\end{equation}
The first term in this expansion comes from a homogeneous false vacuum configuration,
\begin{equation}
Z(\phi_f)=[\det \delta^2 S_E(\phi_f)]^{-1/2} \exp(-S_E(\phi_f)),
\label{fv}
\end{equation}
where $\delta^2 S_E$ denotes the second variation of the Euclidean action with respect to $\phi$. The second term in (\ref{gas1}) comes from the bounce configuration,
\begin{equation}
Z(\phi_b)= \frac{i}{2}\Omega T J \, | {\det}' \delta^2 S_E(\phi_b) |^{-\frac{1}{2}} \exp(-S_E(\phi_b))
\label{bb}
\end{equation}
and higher contributions come from multi-bounce configurations.

In the evaluation of (\ref{bb}) one takes into account that, around the bounce configuration, $\delta^2 S_E$ has one negative mode and four zero modes (the prime on the 
determinant denotes that the zero modes  are excluded). The treatment of the negative mode gives the factor of $i/2$ and the absolute value of the corresponding eigenvalue in the determinant.
 The zero modes can be treated by changing to collective coordinates, which gives a Jacobean $J$, and the integration over these coordinates gives a factor of $\Omega T$, the total physical volume of the system.

The integration over the zero modes can be better illustrated in the case of the $0+1$-dimensional case of a simple quantum-mechanical system with the Euclidean action
\begin{equation}
S^{(1)}(\phi) =\int_{-T/2}^{T/2} dt_E \, \left(\frac{1}{2}(\partial_{t}\phi)^2 + U(\phi)\right),
\end{equation}
with a similar potential that admits a (one-dimensional) bounce solution, $\phi^{(1)}_{b}(t)$, to the Euclidean equation
$\partial_{t}^2\phi = d U/ d \phi$. Because of translational invariance
there is a single zero mode,
\begin{equation}
\phi_{s}=\phi^{(1)}_{b}(t-s)
\end{equation}
is a one-parameter set of solutions with the same action, and
$\phi_0=\partial_{t}\phi_{s}/\sqrt{S^{(1)}_{b}}$, where $S^{(1)}_{b}$ is the action for $\phi^{(1)}_{b}$, is the corresponding zero mode.

The corresponding path integral can be evaluated in the semiclassical approximation around the saddle point \cite{zz1}, after writing $\phi(t)=\phi_{s}+\psi$, expanding $\psi=\sum \alpha_n \psi_n$, in eigenfunctions of $Q=\delta^2 S^{(1)}/\delta^2\phi$, such that $Q\psi_n=\lambda_n\psi_n$, and inserting the identity
\begin{equation}
1=\Delta(\phi)\int ds \, \delta[<\phi_0,\phi-\phi_{s}>],
\label{zm1}
\end{equation}
where $<f,g>=\int f(t)g(t)dt$. The action to second order is $S= S^{(1)}_{b}+\frac{1}{2}\sum\lambda_n\alpha_n^2$, and the divergence of the integration over the zero mode is taken care of by the delta function, resulting in an overall factor of $T$, the ``volume'' of the system, and a Jacobean factor, $J=\Delta(\phi)\approx\Delta(\phi_{s})$ in the semiclassical approximation.

Going back to the $3+1$-dimensional case, one continues with the higher contributions to (\ref{gas1}), which come from multi-instanton configurations: for example, for the two-bounce configuration, dividing the total space-time volume into two volumes $V_1$ and $V_2$, each containing one instanton, one obtains \cite{arnold1}
\begin{eqnarray}
Z_2=\int [d\phi] e^{-S_{2-inst}}&=&\frac{1}{2!}\int[d\phi]_{V_1}\int[d\phi]_{V_2} e^{-\int_{V_1}L_E}e^{-\int_{V_2}L_E}\\ \nonumber
                            &=&\frac{1}{2!}\frac{\int[d\phi]_V e^{-S_1}}{\int[d\phi]_{V_2}e^{-S_0}}
                                           \frac{\int[d\phi]_V e^{-S_1}}{\int[d\phi]_{V_1}e^{-S_0}}\\ \nonumber
                            &=&\frac{1}{2!} Z_f \left(\frac{Z_b}{Z_f}\right)^2,
\label{gas2}
\end{eqnarray}
and
\begin{eqnarray}
Z(T)&=&Z_f + Z_f \frac{Z_b}{Z_f} + Z_f \frac{1}{2!}\left(\frac{Z_b}{Z_f}\right)^2 +\ldots \\ \nonumber
    &=&Z_f \exp{\left(\frac{Z_b}{Z_f}\right)}.
\label{gas3}
\end{eqnarray}

From (\ref{fv}, \ref{bb}) we get
\begin{equation}
\frac{Z_b}{Z_f} =\frac{i}{2} \Omega T A e^{-B}
\label{gas4}
\end{equation}
where $A$ denotes the Jacobian and determinant factors and
\begin{equation}
B=S(\phi_b)-S(\phi_f),
\end{equation}
and, finally, from (\ref{gamma}) we have
\begin{equation}
\Gamma_{\rm flat} =A \, e^{-B}.
\end{equation}
 From the normalization of the four zero modes 
(each one proportional to $\partial_{\mu}\phi_b$)
we get 
a factor of $\sqrt{B/2\pi}$,
to get the Jacobian, $J=(B/2\pi)^2$,
and the full expression for the prefactor is
\begin{equation}
A=\left( \frac{B}{2\pi}\right)^2
\left| \frac{\det'\delta^2 S_E(\phi_b)}{\det \delta^2 S_E(\phi_f)} \right|^{-\frac{1}{2}}.
\label{prefactor1}
\end{equation}
 As a final note one should mention that the dilute instanton gas approximation that was used when considering widely separated instantons in (\ref{gas2}) can be justified when $B >> 1$.

\section{Curved space-time contributions to the decay rate}

In order to get some more quantitative estimates for the expressions and the corrections involved,
I will describe in this Section the curved space-time expressions and compare them to the flat space-time results,
with the approximations used.

When extending the previous results to include the gravitational effects on the false vacuum decay rate,
 one assumes again $O(4)$ symmetry (which has not been rigorously proven in this case) with the metric
\begin{equation}
ds^2 = d\tau^2 + \rho(\tau)^2 d\Omega_3^2,
\label{metric}
\end{equation}
and then the Euclidean action, including gravity, becomes
\begin{equation}
S_E = 2 \pi^2 \int d\tau\left( \rho^3 \left[\frac{1}{2}\dot{\phi}^2 + U(\phi)\right]
                     +\frac{3}{8\pi G}(\rho\ddot{\rho}+\rho\dot{\rho}^2 -\rho)\right),
\end{equation}
(dots will denote derivatives with respect to $\tau$)
for which the Euclidean field equations are:
\begin{equation}
\dot{\rho}^2 -1 = \frac{8\pi G}{3} \rho^2\left[\frac{1}{2}\dot{\phi}^2 - U(\phi)\right],
\label{eq1}
\end{equation}
\begin{equation}
\ddot{\phi} + \frac{3\dot{\rho}}{\rho}\dot{\phi} = \frac{dU}{d\phi}.
\label{eq2}
\end{equation}

I will work in the fixed background approximation where
\begin{equation}
U(\phi)=U_0 + \tilde{U}(\phi)
\end{equation}
with $| \tilde{U}(\phi) |<< U_0 = U(\phi_f)$.
In order to get some more quantitative statements,
I will also consider the extension of the thin-wall approximation from flat space-time,
where $\epsilon=\tilde{U}(\phi_f)-\tilde{U}(\phi_t)$ is small compared to the height of the 
barrier of $\tilde{U}(\phi)$.
It is assumed that $\tilde{U}$ involves a mass scale, $v$, and a
small, dimensionless coupling, $\lambda$.
As before, $\phi_t$ and $\phi_f$ are, respectively, the absolute minimum (true vacuum) and the relative minimum (metastable vacuum) of the potential
and are both of order $v$.
The relevant flat space-time quantities are given by
\begin{equation}
R_b=\frac{3\sigma}{\epsilon}
\label{flatR}
\end{equation}
\begin{equation}
B=\frac{27 \pi^2 \sigma^4}{2 \epsilon^3}
\label{flatB}
\end{equation}
with $\sigma=\int_{\phi_f}^{\phi_t} d \phi \sqrt{2 \tilde{U}(\phi)}$
the surface tension of the bubble.
Generally, $B$ is of order $1/\lambda$, however, the thin wall approximation involves two or more dimensionless parameters \cite{linde1} (for example, quartic and cubic couplings).
For the order-of magnitude statements of this Section, I will denote collectively
 a small dimensionless function of the couplings
by $\lambda$ (some explicit expressions for flat space-time
are given in \cite{linde1}).

With gravity included
(in the fixed background approximation)
 one has a Euclidean de~Sitter background spacetime which is topologically a four-sphere with
\begin{equation}
\rho(\tau) = \frac{1}{H} \sin (H\tau)
\label{ds1}
\end{equation}
where
\begin{equation}
H^2 = \frac{8 \pi G}{3} U_0\equiv\frac{\kappa\,U_0}{3}\equiv\frac{8 \pi \, U_0}{3 M_P^2}
\end{equation}
and the coordinate $\tau$ extends from $0$, on the ``south pole'', to $\frac{\pi}{H}$, on the ``north pole''  of the sphere.

The bounce solution to (\ref{eq2}),
\begin{equation}
\ddot{\phi} + 3H \cot (H\tau) \dot{\phi} = \frac{dU}{d\phi}.
\label{b1}
\end{equation}
 takes values close to $\phi_t$ at the ``south pole'', $\tau\approx 0$, and falls rapidly to $\phi_f$,
in the thin wall approximation,
with a radius $\tilde{R}_b$,
 at larger values of $\tau$, satisfying the boundary conditions $\dot{\phi}=0$ at $\tau=0$ and $\tau=\frac{\pi}{H}$.

Provided that the bounce radius, $\tilde{R_b}$, is much smaller than $\frac{\pi}{H}$ one can justify the dilute instanton gas approximation described in the previous section and proceed to derive the bubble nucleation rate,
$\Gamma_{\rm curved} =\tilde{A} e^{-\tilde{B}}$
with
\begin{equation}
\tilde{B}=2\pi^2 \int d\tau \rho^3 \left[ \frac{1}{2}\dot{\phi}^2 + \tilde{U}(\phi)\right].
\label{exp1}
\end{equation}

With the thin wall and fixed background approximations used, the relevant quantities become \cite{park}
\begin{equation}
\tilde{R_b}^2=\frac{R_b^2}{1+(HR_b)^2},
\end{equation}
\begin{equation}
\tilde{B}=\frac{B}{1+(H R_b)^2/2},
\end{equation}
and the prefactor (with ``frozen'' gravity) becomes
\begin{equation}
\tilde{A}=\left( \frac{\tilde{B}}{2\pi}\right)^2
\left| \frac{\det'\delta^2 S_E(\phi_b)}{\det \delta^2 S_E(\phi_f)} \right|^{-\frac{1}{2}}.
\label{prefactor1}
\end{equation}
The first factor here, $(\tilde{B}/2\pi)^2$, comes from the four zero modes of the bounce on the four-sphere
(it is not obvious in the coordinates used but one can easily see the generalization of the flat space-time
zero modes and their normalization using conformally flat coordinates in our approximations).
The determinants involve the Laplacian on the four-sphere for the scalar field only
(the gravitational background of the de~Sitter space-time is considered fixed)
\begin{equation}
\delta^2 S_E = - H^2 \Box_y  + U''(\phi),
\end{equation}
where $y=H \tau$ and $\Box_y$ is the dimensionless Laplacian on a four-sphere
of unit radius
(primes on $U$ denote derivatives with respect to $\phi$).
If the regularized eigenvalues of the determinants in flat space-time
are denoted generically by $\omega\sim\lambda v^2$, their curved space-time
counderparts
 become $\tilde{\omega}\sim\omega(1+ c_d (H R_b)^2)$
with $c_d$ a numerical constant of order unity.
Since the determinant around the bounce has four less eigenvalues (the zero modes are excluded) the
power counting arguments for the full expression for the bubble nucleation rate
give
\begin{equation}
\Gamma_{\rm curved} \sim \frac{B^2}{(1+(HR_b)^2/2)^2}
(\lambda v^2)^2  (1+c_d (HR_b)^2)^2
e^{-\frac{B}{(1+(HR_b)^2/2)}}.
\label{g2}
\end{equation}
$B$ and $R_b$ are the flat space-time quantities (\ref{flatR}), (\ref{flatB})
and,
as mentioned before, in the thin wall approximation, there are more dimensionless 
parameters besides $\lambda$ to be included in these expressions \cite{linde1}. I also note that, in our approximations,
after differentiating (\ref{b1}) with respect to $\tau$, one sees that there is
a negative mode \cite{neg} for the determinant around the bounce in de~Sitter space-time, with an eigenvalue of order
$-3/\tilde{R_b}^2$, which agrees with the power counting arguments given before.

\section{The bounce and the instanton gas in de~Sitter space-time}

The general expression for the tunneling rate,
possibly when some of the approximations used here are relaxed,
\begin{equation}
\Gamma_{\rm curved} = \tilde{A}(\lambda, H, v) e^{-\tilde{B}(\lambda, H, v)},
\end{equation}
considered
as a function of the parameters of the theory,
may contain some overall exponentially small terms
($e^{-B}$ in our case, as described in the previous Section),
but also contains terms of order unity, and it is possible that, in a landscape or multiverse scenario,
it has a global or local maximum at specific values of the physical parameters
(masses, couplings, and especially the cosmological constant).

In fact, it is not simply the expression for the tunneling rate that is expected to have an extremum,
as much as quantities that show the rate of progress of the cosmological phase transition \cite{ww}
but in any case, $\Gamma$ is one of
the main inputs necessary for the calculation of these quantities.

It should be obvious that, even with the approximations used here,
 in curved space-time, there is a multitude of possibly comparable contributions
originating from different physical  effects \cite{ejw1, k1}. In our case,
since the Euclidean de~Sitter four-sphere has a finite size, $\tau$ extends up to the de~Sitter horizon
\begin{equation}
R_{dS}=\frac{\pi}{H},
\label{hor}
\end{equation}
and there are always contributions from the interactions between instantons as they are distributed on the four-sphere, especially when their radius grows larger.

Here I will estimate the first corrections to $\Gamma_{\rm curved}$ that appear because of this effect. In order to get a quantitative estimate I will assume that we are still inside the limits of the previous approximations (in particular, the condition $B >> 1$ also has to hold, as in the flat case, and the thin wall and fixed background limits are also assumed) and obtain, therefore, a correction to the pre-exponential factor for $\Gamma$, which is of interest, however, since it depends on the parameters of the model and especially on the cosmological constant of the background de~Sitter space-time.

A first estimate of the corrections involved comes from considering the two-instanton interaction, as it appears in (\ref{gas2}) and its generalization to the Euclidean de~Sitter spacetime. The method involved is similar to the calculation of soliton interactions in \cite{manton} and a basic reason that it can be used here in order to get quantitative results is that the instanton interaction in de~Sitter spacetime turns out to be repulsive, essentially because of the cosmological expansion rate, and unlike most scalar soliton and instanton interactions in quantum field theory. The two-instanton configuration that gives the greatest contribution to the saddle-point evaluation of (\ref{gas2}) will consist, therefore, of two bounce solutions symmetrically on the ``north pole'' and ``south pole''
of the de~Sitter four-sphere.

In order to justify the previous statements we consider, accordingly, a two-instanton configuration of the form
\begin{equation}
\phi_s(\tau) = \tilde{\phi}_b(\tau) +\tilde{\phi}_b(s-\tau) +\phi_f
\label{2inst}
\end{equation}
where
\begin{equation}
\tilde{\phi}=\phi -\phi_f
\end{equation}
and $\phi_b$ a solution to the bounce equation (\ref{b1}).
When the two-instanton separation $s$ is very close to $\frac{\pi}{H}$ the previous configuration is an approximate solution of the Euclidean field equations and one can estimate their interaction as a function of $s$. As in \cite{manton}, the interaction will come from the overlap of the ``tails'' of the two widely separated instantons, that is around the ``equator'' $\tau =\frac{\pi}{2H}$. One needs, therefore, the asymptotic behavior of the bounce solution to (\ref{b1}).
First we write around the false vacuum
\begin{equation}
U(\phi) = U_0 +\frac{\beta H^2}{2} (\phi -\phi_f)^2
\end{equation}
with
\begin{equation}
\beta = \frac{U''(\phi_f)}{H^2}.
\end{equation}
After setting $x=\tau -\frac{\pi}{2H}$, writing $\cot(H\tau)\approx -Hx$ for small $x$,
changing to $\tilde{\phi} = u(x) e^{\frac{3 H^2}{4} x^2}$ and neglecting terms of higher order in $x$ in the final expressions, we find the asymptotic behavior for $\tilde{\phi} = \phi - \phi_f$ around $\tau \approx \frac{\pi}{2H}$:
\begin{equation}
\tilde{\phi}\approx c \, e^{-H (\beta -\frac{3}{2})^{1/2}\,\tau}
\end{equation}
with $c$ a constant of order $v$.

Now one can estimate the two-instanton interaction $\delta B$ as the difference between the value for $B$ of the configuration $\phi_s$ and twice the value for a single bounce,
\begin{equation}
B(\phi_s) = 2 B + \delta B,
\end{equation}
and, using the previous expressions, we find
\begin{eqnarray}
\delta B &=& 2\pi^2 \int d\tau \rho^3 \left[\dot{\tilde{\phi}}_b(\tau) \dot{\tilde{\phi}}_b(s-\tau)
                            + \beta H^2 \tilde{\phi}_b(\tau) \tilde{\phi}_b(s-\tau) \right] \\ \nonumber
  &=&\left(2 \pi^2 c^2 \int_{eq} d\tau \frac{1}{H^3}\sin^3 (H\tau)\right) \frac{3}{2}H^2 e^{-H(\beta -\frac{3}{2})^{1/2}s}   \\ \nonumber
  &=& 2\pi^2 c^2 \frac{4}{3 H^4} \frac{3}{2}H^2 e^{-H (\beta -\frac{3}{2})^{1/2} \, s}.
\end{eqnarray}
Here,
the integration inside the parenthesis in the second line of the previous expression is done
around the ``equator'' $\tau\approx \frac{\pi}{2H}$, since, however, the integrand falls to zero away from this value, I have extended the region of integration to the entire range, the difference being a numerical factor of order unity that can, in any case,      be absorbed in $c$.

One can see immediately from this expression that the effective instanton interaction in the background de~Sitter spacetime is repulsive, and, consequently, that the path integral is dominated by growing values of $s \approx \frac{\pi}{H}$.

If this instanton interaction term did not exist, one would proceed from expressions like (\ref{gas2}) to derive
$\Gamma_{CDL}$ as before, the integration over $s$, from $s=0$ to $s=\frac{\pi}{H}$, corresponding to a collective coordinate contributing to the total volume factor in (\ref{gas4}).
Now, however, this integration will give a different factor, leading to an additional contribution to the pre-exponential factor for the false vacuum decay rate.
In the two-instanton interaction, the factor $Z_2$ in (\ref{gas2}), after treating (\ref{2inst})
in the manner of (\ref{zm1}),  gets modified to
\begin{equation}
Z_2 =\frac{1}{2!}\alpha \, Z_f  \left(\frac{Z_b}{Z_f}\right)^2 ,
\end{equation}
with
\begin{equation}
\alpha = \frac{\int_0^{\pi/H} ds \, e^{-\delta B}}{\pi/H}
\end{equation}
and
in the general $n$-instanton contribution to (\ref{gas2}) one may consider the nearest neighbor interaction and integrate out each instanton to get the general factor
\begin{equation}
Z_n = \frac{1}{n!} \left(\frac{Z_1}{Z_0}\right)^n  \alpha^{n-1}.
\label{gas5}
\end{equation}
Overall this leads to a correction to the pre-exponential factor
of the tunneling rate given in the previous Section
\begin{equation}
\Gamma_{\rm curved} =\alpha\,\tilde{A}\, e^{-\tilde{B}}
\end{equation}
with
\begin{equation}
\alpha(\lambda, H, v)=\frac{1}{\pi}\int_0^{\pi} dx \exp(-a_1 e^{-a_2 x})
\label{result1}
\end{equation}
where $a_1 =\frac{4 \pi^2  c^2}{H^2}$ and $a_2=\sqrt{\beta -\frac{3}{2}}$.
When $\frac{H}{v}<<1$ this has the asymptotic expansion $\alpha \approx 1 -\frac{\ln a_1}{\pi a_2} -\frac{\gamma}{\pi a_2}$, where $\gamma = 0.57721...$ is the Euler-Mascheroni constant.
It is of the same order of magnitude as the other contributions to the prefactor that were described in the previous Section;
however,
since it depends on the same parameters of the theory, it should be included when considering the behaviour of the 
bubble nucleation rate and the overall progress of a cosmological phase transition.

\section{Comments}

In summary, using the thin-wall, dilute instanton gas approximations of \cite{cdl} and the fixed-background approximation as described before, I derived the first corrections to the
pre-exponential factor
of the Coleman-De~Luccia vacuum decay rate that originate from the instanton interactions inside the finite horizon of the background de~Sitter spacetime.

It should be stressed again that the general configuration considered in (\ref{2inst}) is only an approximate solution to the Euclidean field equations, when the instanton separation is very close to the horizon value $s = \pi/H$; because, however, of the repulsive nature of the instanton interaction, the integrals considered are dominated by these values (this is reflected by the fact that the leading term of the above expressions for the corrections is unity). 
 It would be interesting to see whether some of the approximations used can be relaxed; it is possible that this may be true for the thin wall approximation, since the instanton interaction comes mainly from the overlap of the ``tails'' of two bounce configurations. It would also be useful, from the phenomenological point of view, to investigate the limits of the fixed-background approximation and extend the results to include the case of a negative cosmological constant and anti-de~Sitter space-times.
Hopefully these issues, as well as the application of these results to models of the cosmological landscape will be the subject of future work.
        
\vspace{0.5in}

\centerline{\large\bf  Acknowledgements}
\noindent
This work was completed at the National Technical University of Athens. I would like to thank the people of the Physics Department for their support 
and especially Kostas Farakos for many useful discussions.

\end{document}